\documentclass{ar2e}
\begin{document}
\input epsf.tex    

\input psfig.sty

\newcommand{\lsim}{\lesssim}
\newcommand{\gsim}{\gtrsim}
\newcommand{\tr}{{\rm Tr}}

\jname{Annu. Rev. Nucl. Part. Sci.}
\jyear{2010}

\title{Supersymmetry Breaking \\ and Gauge Mediation}

\markboth{Ryuichiro Kitano,
Hirosi Ooguri and
Yutaka Ookouchi}{supersymmetry breaking and gauge mediation}

\author{
Ryuichiro Kitano$^{1}$,
Hirosi Ooguri$^{2,3}$ and
Yutaka Ookouchi$^4$
\affiliation{\it 
${}^1$Department of Physics, Tohoku University, Sendai 980-8578, Japan\\ 
${}^2$California Institute of Technology, Pasadena, CA 91125, USA\\
${}^3$Institute for the Physics and Mathematics of the Universe, \\ 
University of Tokyo, Kashiwa, Chiba 277-8582, Japan\\
${}^4$Perimeter Institute for Theoretical Physics, ON N2L2Y5, Canada}}


\begin{abstract}
We review recent works on supersymmetry breaking and gauge mediation. We
survey our current understanding of dynamical supersymmetry breaking
mechanisms and describe new model building tools using duality,
meta-stability, and stringy construction.  We discuss phenomenological
constraints and their solutions, paying attentions to issues with
gaugino masses and electroweak symmetry breaking.

\end{abstract}

\maketitle

\section{Introduction}

Supersymmetry has been playing important roles in modern particle
physics even though there has been no direct experimental evidence for its
existence in Nature.
It is a hypothetical space-time symmetry which transforms
bosonic states into fermionic ones and vise-versa, but this funny
symmetry has nice features, such as vacuum stability ($E \geq 0$) and 
mild ultraviolet behavior of theories,
$i.e.$, a restricted form of divergences.
Many theorists expect that
supersymmetry is an essential ingredient for the ultimate unified
theory of elementary particles including gravity, perhaps
the string theory.

Not only as a possible symmetry of Nature, supersymmetry have provided
examples of (partly) calculable strongly coupled theories. Of the most
amazing is the discoveries of strong/weak dualities among supersymmetric
theories, such as dualities in the ${\cal N}=4$ supersymmetric
Yang-Mills theory~\cite{OMI,OMV,OMII,OMIV,OMIII}, the Seiberg-Witten
theory \cite{Seiberg:1994rs}, the Seiberg dualities
\cite{Seiberg:1994pq, Seiberg:1994bz} in QCD-like supersymmetric gauge
theories, and the gauge/gravity correspondence \cite{Maldacena}. These
dualities boosted our understanding of non-perturbative physics.

With or without the connection to the theory of quantum gravity, particle
theorists have applied supersymmetry to the Standard Model of Particle
Physics. Their main motivation is to protect the Higgs potential against
quantum corrections. Because of very weak ultraviolet
divergences, one can naturally
push the cut-off scale to an arbitrary high energy scale, such as the
Planck scale ($\sim 10^{18}$~GeV).

Once one postpones the cut-off scale until, say, the Planck
scale, it offers a wonderful arena for model-builders. One can build
calculable field theoretical models of high-scale or high-temperature
phenomena such as the inflation, baryogenesis, neutrino masses, fermion mass
hierarchies, grand unification etc., without worrying about naturalness
problems caused by large quantum corrections.

The successful unification of the coupling constants in the 
minimal supersymmetric extension of the Standard
Model~\cite{Dimopoulos:1981zb, Sakai:1981gr,
Dimopoulos:1981yj} has attracted many theorists 
to the supersymmetric world.
Moreover, the lightest supersymmetric particle is
a strong candidate for dark matter of the universe.
See \cite{Jungman:1995df} for a review.
Because of this beautiful and successful framework, there have been many
studies on supersymmetry searches at high-energy colliders such as LEP,
Tevatron, LHC, and ILC, and also through dark matter
detections~\cite{Amsler:2008zzb}.

However, there is one missing piece, that is supersymmetry breaking
mechanism. 
We need mass splittings between bosons and fermions because of 
experimental constraints. Satisfying the experimental bounds themselves
is not hard. One can assume a supersymmetry breaking sector which
couples to the minimal supersymmetric standard model (MSSM) in such 
a way that large enough mass splittings are obtained.
A non-trivial constraint on model building comes from consideration of
the Higgs sector.
Naturalness of electroweak symmetry breaking together with the
experimental bounds require that all the superpartners have masses of
${\cal O}(100~{\rm GeV})$.

In this article, we review recent developments towards a viable
supersymmetric model of particle physics.
We start with discussion of an early attempt at supersymmetric model
building by Dine, Fischler, and Srednicki \cite{DFS}, in which the
concepts of dynamical supersymmetry breaking, (direct) gauge mediation
and their connection to the electroweak symmetry breaking have been
discussed.
Even though the specific model presented in the paper has been found 
to be incomplete
due to progress in supersymmetric gauge theories, their
scenario remains one of the most attractive and elegant.
We then jump to recent progress and current understanding of those
topics, including the discovery of meta-stable vacua in the
supersymmetric QCD (SQCD) by Intriligator, Seiberg and Shih
(ISS)~\cite{ISS}, new formulations of gauge mediation and related
topics, direct gauge mediation models and discussion on the gaugino
masses, and connections to string theory.
We close the discussion with open problems associated with electroweak
symmetry breaking\footnote{There are numbers of topics which we do not
cover in this review article. Especially, we do not discuss cosmological
or astrophysical constraints on supersymmetric models. We refer to, {\it
e.g.}, \cite{Moroi:1993mb, Banks:1993en, Bagger:1994hh} for constraints
relevant for dynamical supersymmetry breaking and gauge mediation.}.

\section{Prototype scenario of low energy supersymmetry}

Witten in 1981 started asking a question of whether supersymmetry
breaking happens dynamically. In ~\cite{DSBWitten}, he proposed a
natural framework for electroweak physics, ``supersymmetric
technicolor,'' where a technicolor force dynamically breaks both
supersymmetry and the electroweak $SU(2)\times U(1)$.
Independently, in the same year, Dine, Fischler, and Srednicki~\cite{DFS},
and Dimopoulos and Raby~\cite{DR} proposed concrete models along the similar
line. Especially in \cite{DFS}, a mechanism to generate gaugino and sfermion
masses has been discussed. This mechanism is now called direct gauge
mediation.
We first recall this early attempt in this section.

\begin{figure}
\centerline{\epsfbox{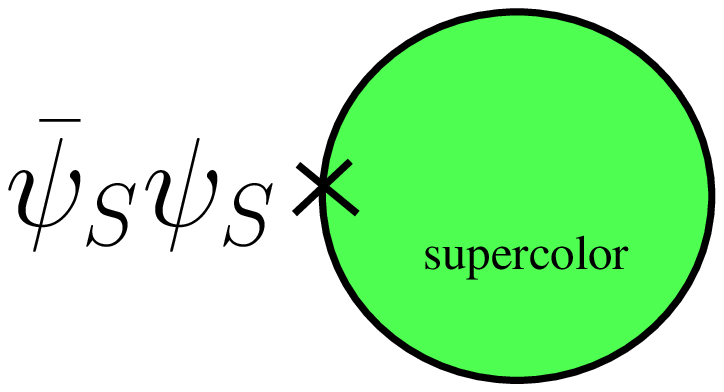}}
\caption{Conceptual picture for dynamical supersymmetry breaking.}
\label{fig:condensation}
\end{figure}

Their picture is quite simple. There is a QCD-like $SU(M)$ gauge theory
which becomes strong at a scale $\Lambda_{\rm SC}$ (SC stands for
supercolor), and it is assumed that supersymmetry is dynamically
broken by a condensation of a pair of the ``superquarks,'' $\psi_S$ and $\bar
\psi_S$, which are fermion components of chiral superfields, $S$ and
$\bar S$ (Figure~\ref{fig:condensation}).  Since a condensation of the
fermion pair, $\langle \bar \psi_S \psi_S \rangle \simeq \Lambda_{\rm
SC}^3$, is a $F$-component of the meson superfield ($M \sim S \bar S$),
supersymmetry will be spontaneously broken if such a condensation forms.

\begin{figure}
\centerline{\epsfbox{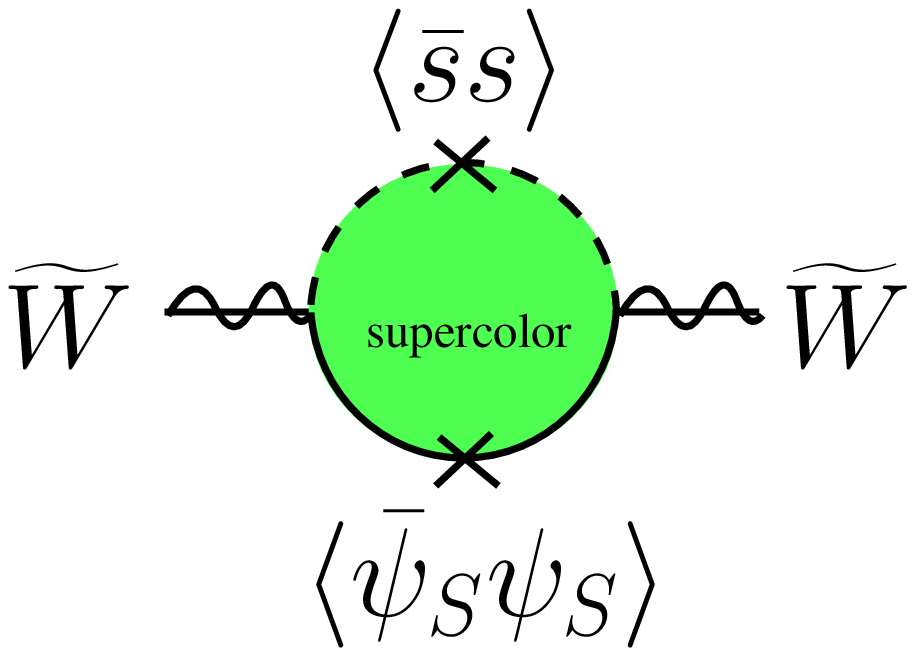}}
\caption{A diagram to contribute gaugino masses.}
\label{fig:gaugino}
\end{figure}

\begin{figure}
\centerline{\epsfbox{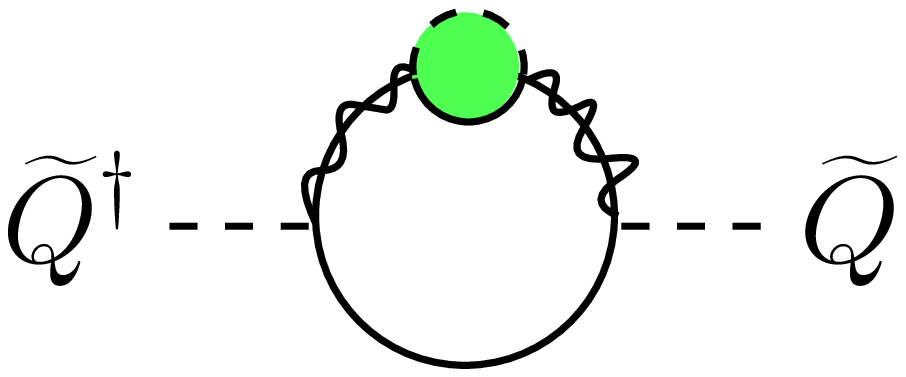}}
\caption{One of the diagrams to contribute sfermion masses.}
\label{fig:sfermion}
\end{figure}

A part of the global symmetry in the sector is identified with the
standard model gauge group by assigning quantum numbers to the
superquarks.  (In the original model only the $SU(2)\times U(1)$ part was
embedded.) The gauginos and sfermions can obtain masses through loop
diagrams involving the standard model gauge interactions
(Figures~\ref{fig:gaugino} and \ref{fig:sfermion}).
To generate non-zero gaugino masses, it is assumed that
there is a boson-pair condensation 
\begin{eqnarray}
 \langle \bar s s \rangle \simeq \Lambda_{\rm SC}^2,
\end{eqnarray}
in addition to the fermion-pair condensation.  Here $\bar s$ and $s$ are
the lowest components of $\bar S$ and $S$, respectively. The presence of
both condensations ensures breakdown of an R-symmetry necessary for
non-vanishing gaugino masses.
In order to obtain ${\cal O}(100)$~GeV gaugino and sfermion masses, the
dynamical scale $\Lambda_{\rm SC}$ was assumed to be ${\cal O}(10)$~TeV
due to the loop factors ($\sim \alpha / 4 \pi$).

\begin{figure}
\centerline{\epsfbox{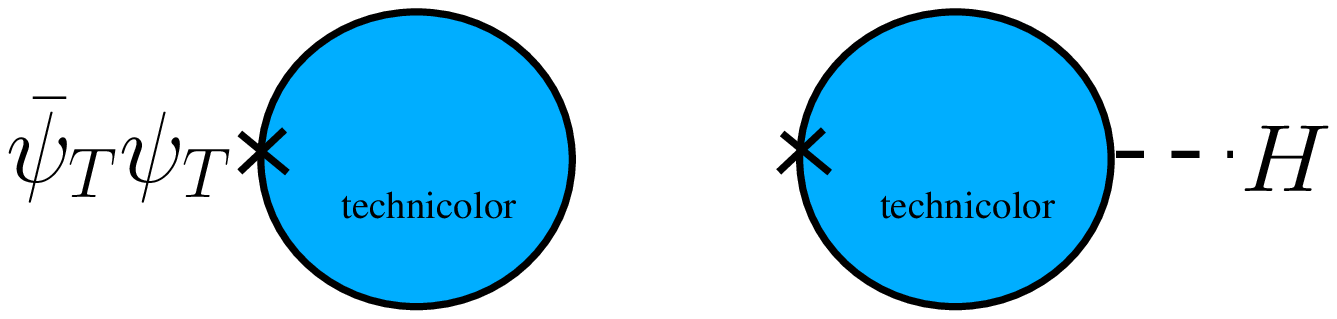}} 
\caption{Pictures for electroweak
symmetry breaking and the VEV of the Higgs field.}  \label{fig:techni}
\end{figure}

\begin{figure}
\centerline{\epsfbox{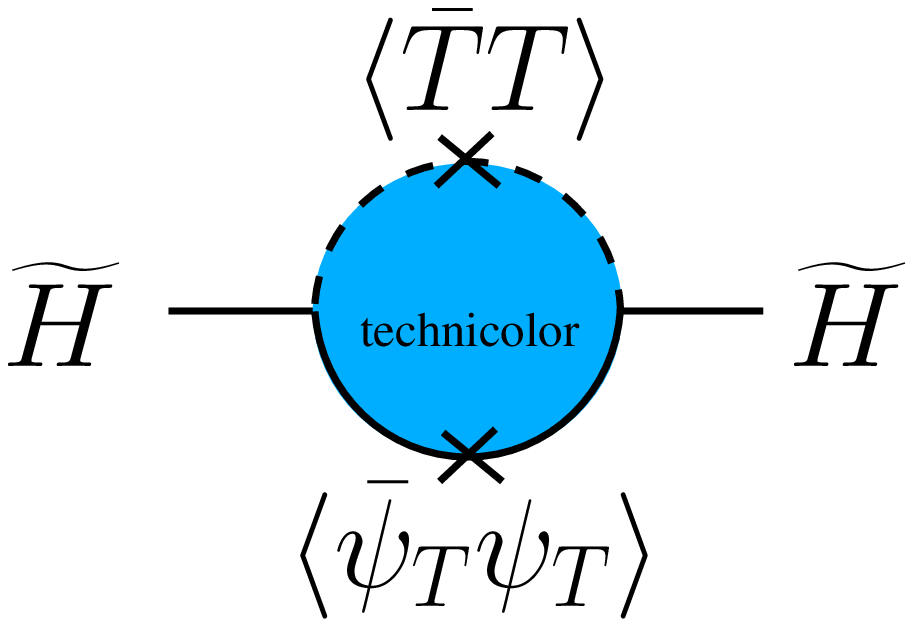}} 
\caption{A diagram to contribute the Higgsino mass.}  \label{fig:higgsino}
\end{figure}

Finally, the electroweak symmetry breaking can happen either through
some other dynamics (technicolor) or the standard Higgs mechanism. 
In both cases, one can introduce elementary Higgs fields to write down
the Yukawa interactions while avoiding the naturalness issue thanks to
supersymmetry.
In \cite{DFS}, a technicolor model is proposed where the electroweak VEV
and the higgsino mass are generated through a strong dynamics at the
${\cal O}(300~{\rm GeV})$ energy scale as illustrated in
Figures~\ref{fig:techni} and \ref{fig:higgsino}. (The possibility of
using the usual Higgs mechanism was also discussed in the concluding
section.)

We can see three essential ingredients in this model: dynamical
supersymmetry breaking with R-symmetry violation, gauge mediation, and
electroweak symmetry breaking through supersymmetry breaking.
In the following, we will review recent progress and current
understanding of those three components.

\section{Supersymmetry breaking}\label{sec:susybreaking}

In our current understanding, the assumption made in \cite{DFS} for
supersymmetry breaking,
\begin{eqnarray}
 \langle \bar \psi_S \psi_S \rangle \simeq \Lambda_{\rm SC}^3,
\label{eq:condense}
\end{eqnarray}
is not valid in supersymmetric QCD theories. It has been shown that the
these theories have stable supersymmetric
vacua~\cite{WittenIndex,NS}. Alternative possibilities have been
considered and successful models for dynamical supersymmetry breaking
have been found, for example chiral gauge theories in
\cite{ADSI,ADSII,ADSIII} and theories with gauge singlet fields in
\cite{IzawaYanagida,IntriligatorThomas}. See
\cite{Izawa:2009mj,Izawa:2009nz} for more recent proposals.

\begin{figure}
\centerline{\epsfbox{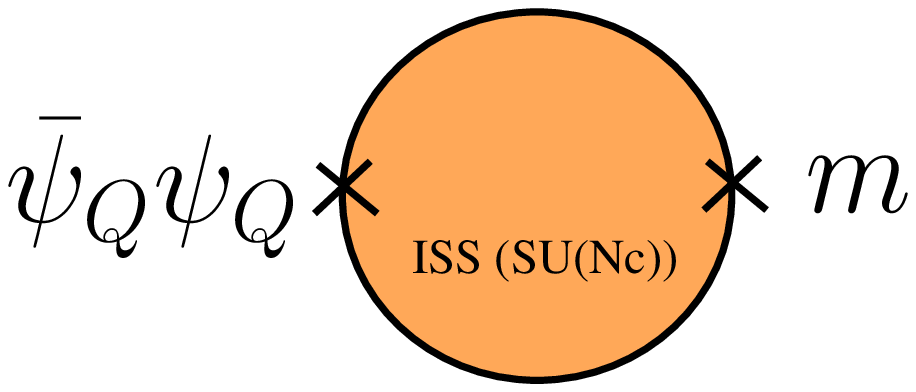}} 
\caption{A picture for dynamical supersymmetry breaking in the ISS
 model.}  
\label{fig:ISS}
\end{figure}

Recently, dynamical supersymmetry breaking in supersymmetric QCD
theories has revived by the work of ISS. They found a meta-stable
supersymmetry breaking vacuum in $SU(N_c)$ supersymmetric gauge theories
with massive $N_f$ flavors for $N_c < N_f < 3N_c/2$.
The presence of the vacuum is established when $m \ll \Lambda$ with $m$
and $\Lambda$ the quark mass and the dynamical scale, respectively.
A non-vanishing fermion-pair condensation is obtained to be
\begin{eqnarray}
 \langle \bar \psi_Q \psi_Q \rangle \sim m \Lambda^2,
\end{eqnarray}
instead of Equation~(\ref{eq:condense}) as illustrated in
Figure~\ref{fig:ISS}.

Let us first discuss the linear sigma models for supersymmetry breaking
which are widely used as a tool to establish existence of supersymmetry
breaking vacuum in strongly coupled theories~\cite{ADSI,ADSII, ADSIII,
CLP, Murayama} and also as effective theories to describe low energy
physics of a variety of supersymmetry breaking models.
We then introduce the ISS model and its connection to the linear
sigma model.
For a recent pedagogical review on supersymmetry breaking, see
\cite{Intriligator:2007cp}.

\subsection{Polonyi and generalized O'Raifeartaigh models}

Supersymmetry is spontaneously broken when the $F$-component of a chiral
superfield $X$ acquires a VEV:
\begin{eqnarray}
 \langle F_X \rangle \neq 0,
\end{eqnarray}
since $2 \langle F_X \rangle = \langle \{ iQ, X \}|_{\theta} \rangle$
which should vanish if $Q|0\rangle = 0$.
It is easy to construct a linear sigma model to describe this phenomenon
using a single chiral superfield $X$. The superpotential is simply,
\begin{eqnarray}
 W = \mu^2 X.
\label{eq:super}
\end{eqnarray}
The $F$-component of $X$ acquires a VEV unless the K{\" a}hler potential
is singular. This is the unique choice of the superpotential for a model
with a single chiral superfield $X$. (There can be a small perturbation
to it if one allows the vacuum to be meta-stable.)
By an appropriate shift of $X$ we can choose the stable point at $X =
0$. In order for this point to be stable, the K{\" a}hler potential
expanded around it is assumed to be of the form:
\begin{eqnarray}
 K = X^\dagger X - {(X^\dagger X)^2 \over \Lambda^2} + \cdots .
\label{eq:kahler}
\end{eqnarray}
Equations~(\ref{eq:super}) and (\ref{eq:kahler}) define the linear sigma
model for supersymmetry breaking, the Polonyi model~\cite{Polo}.

The equation of motion for $F_X$ and the potential minimization of $X$
leads
\begin{eqnarray}
\langle F_X \rangle = \mu^2, \ \ \ \langle X \rangle = 0.
\end{eqnarray}
Therefore, supersymmetry is spontaneously broken. The fermion component
$\psi_X$ remains massless. This is the Goldstino fermion associated with
the spontaneous supersymmetry breaking.
The complex scalar field $X$ obtains a mass:
\begin{eqnarray}
 m_X^2 = {4 \mu^4 \over \Lambda^2 }.
\end{eqnarray}
Up to ${\cal O}(1/\Lambda^2)$ there is no mass splitting between the scalar and
the pseudo-scalar parts due to an unbroken approximate R-symmetry with
$R(X) = 2$.

Although it is a non-renormalizable model, the model serves as the
effective theory (if $\mu^2 \ll \Lambda^2$) for a wide class of
supersymmetry breaking models.
Conversely, one can establish the presence of a supersymmetry breaking
vacuum when a model reduces to the Polonyi model at a point of the field
space by integrating out massive degrees of freedom.  This technique has
been used
in~\cite{ADSI,ADSII,ADSIII,CLP,Murayama,ISS,OOP,MOOP,KitanoOokouchi}.

One can also use the Polonyi model as a hidden sector whose
supersymmetry breaking effects are communicated to the MSSM sector
through some interactions such as gravity or gauge interactions.
Especially in
\cite{SweetspotI,SweetspotII,MurayamaNomura,MurayamaNomuraII}, simple
models of gauge mediation have been constructed by attaching the
messenger fields to the above linear sigma model.
They belong to the indirect type of gauge mediation in
contrast to the direct one depicted in Figure~\ref{fig:gaugino}
and discussed later.

The O'Raifeartaigh model~\cite{ORaifeartaigh:1975pr} and its
generalization are alternative simple theories for supersymmetry
breaking.
(They reduce to the Polonyi model when massive fields are much heavier
than the size of supersymmetry breaking.)
The generalized O'Raifeartaigh model is defined by a set of chiral
superfields ($\Phi_i$) and a superpotential up to cubic terms (the
Wess-Zumino model~\cite{Wess:1973kz}) where
\begin{eqnarray}
 {\partial W \over \partial \Phi_i} = 0
\end{eqnarray}
cannot be solved simultaneously. The chiral superfields $\Phi_i$ have
the canonical K{\" a}hler potential, $K = \Phi_i^\dagger \Phi_i$.
These models have received renewed attention recently because many
dynamical models (including the ISS model) reduce to it at low energy.
One of the basic property common to all the O'Raifeartaigh models is
existence of tree-level flat directions, known as pseudo-moduli,
emanating from any local supersymmetry breaking vacuum~\cite{Ray}. Such
pseudo-moduli receive quantum corrections and a potential along the flat
direction is generated. 
%

\subsection{ISS model}

It has been known from the studies by Seiberg~\cite{Seiberg:1994bz} that
supersymmetric QCD theories with the quark mass $m$ has stable
supersymmetric vacua at
\begin{eqnarray}
 \langle M \rangle = \langle Q \bar Q \rangle \sim
\Lambda^{(3 N_c - N_f)/N_c} m^{(N_f - N_c)/N_c},
\label{eq:qcd}
\end{eqnarray}
where $N_c$ and $N_f$ are numbers of color and flavor, respectively.
This fact appeared to be a no-go theorem for dynamical supersymmetry
breaking in these theories~\cite{WittenIndex}.

Recently, it was shown by ISS that there can be a meta-stable
supersymmetry breaking vacuum far away from the supersymmetric vacua.
The model they considered is a supersymmetric $SU(N_c)$ theory with
$N_f$ fundamental and anti-fundamental chiral superfields $Q_i$ and
$\bar Q_i$ ($i=1,\cdots,N_f$). The superpotential is
\begin{eqnarray}
 W = m Q_i \bar Q_i,
\label{eq:mass}
\end{eqnarray}
where the color and flavor indices are contracted. This is an
asymptotically free theory and becomes strongly coupled in the
IR. Therefore, the analysis in the electric picture (description in
terms of quarks and gluons) is difficult near the origin of the meson
space $M \sim Q \bar Q$.
By the power of duality, the IR physics of the model is described
by the dual magnetic theory, which is weakly coupled 
for $N_c < N_f < 3N_c/2$~\cite{Seiberg:1994pq}. 
Perturbative calculations in the magnetic theory are reliable near the
origin of the meson field space.

The dual gauge theory is $SU(N_f - N_c)$ with $N_f$ dual quarks, $q$ and
$\bar q$, and the gauge singlet meson fields $M$. The superpotential is
\begin{eqnarray}
 W_{\rm mag.} = h \mu^2 M_{ii} - h  q_i M_{ij} \bar q_j,
\end{eqnarray}
where $h$ is a coupling constant of ${\cal O}(1)$ and the term $h\mu^2
M$ corresponds to the quark mass terms in Equation~(\ref{eq:mass}).  The
coefficient $h \mu^2$ is naturally of ${\cal O}(m \Lambda)$.  Again, $i$
and $j$ are the $SU(N_f)$ flavor indices.
At the point where $\langle M_{ij} \rangle=0$ and $\langle q_i \rangle =
\langle \bar q_i \rangle \neq 0$ and the dual gauge group is completely
broken, the flavor $SU(N_f)$ symmetry is broken down to $SU(N_f -N_c)
\times SU(N_c)$.
The effective superpotential at that point has the form of the
O'Raifeartaigh model\footnote{For simplicity we omitted other
pseudo-moduli in the model. For detail, see \cite{ISS}.}:
$${ W= h \tr \left( \mu^2 X - X \rho \tilde{\rho} - \mu \rho \tilde{Z} -
\mu Z\tilde{\rho} \right),  }$$ 
where $X$ is the pseudo-moduli field, which is a $N_c \times N_c$ part
of the meson $M$, and $\rho$, $\tilde \rho$, $Z$ and $\tilde Z$ are
massive fields.  Once we integrate out these massive fields, the
Coleman-Weinberg potential is generated, which stabilize the
pseudo-moduli at the origin \cite{ISS}.
The $F$-component of the $X$ field $(\sim \bar \psi_Q \psi_Q)$ gets a
VEV triggered by an explicit breaking of chiral symmetry as in
Figure~\ref{fig:ISS}.
Although the detailed structure is different, the dynamical
supersymmetry breaking through a fermion-pair condensation assumed
in~\cite{DSBWitten,DFS,DR} is revived in the meta-stable vacua of
supersymmetric QCD models.

The tunneling rate into the true supersymmetric vacuum in
Equation~(\ref{eq:qcd}) is exponentially suppressed if $m \ll \Lambda$
since the supersymmetric vacuum is located far from the meta-stable
vacuum, in comparison with the height of the potential along the meson
direction $V^{1/4} \sim (m^2 M^2)^{1/4}$.

\section{Gauge Mediation}

Inspired by the simplicity and genericity of meta-stable supersymmetry
breaking, model building of gauge mediation have been invigorated
lately\footnote{See \cite{Giudice:1998bp} for a review on earlier works
on gauge mediation.}. The diagram in Figure~\ref{fig:gaugino} has been
reconsidered with new knowledge of supersymmetry breaking, and also been
reformulated by using current correlators.
In this section, we review techniques to calculate the gaugino and
sfermion masses.

\subsection{Analytic Continuation into Superspace \label{analytic}}

One can perform explicit calculations of gaugino and sfermion masses
when the supercolor boxes in Figures~\ref{fig:gaugino} and
\ref{fig:sfermion} are loops of weakly coupled fields, called the
messenger fields. Such a class of models are first considered in
\cite{Dine:1981gu, AlvarezGaume:1981wy, Nappi:1982hm} and revived in mid
'90s by \cite{DineNelson,DineNelsonShirman,DNY} where explicit
computations of gaugino and sfermion masses are performed. The highly
predictive feature of gauge mediation is demonstrated; the gaugino and
sfermion masses are functions only of their quantum numbers to a good
approximation.

In \cite{GR,AGLR}, a powerful method was developed to compute multi-loop
quantities from one-loop running data. The results follow from imposing
constraints due to holomorphy in a spurion $X=X_0+\theta^2 F$ on the
effective action.
One can use the method as the leading order calculation in the $F/X_0^2$
expansion; higher order terms in $F$ arise from terms involving
super-derivatives, which are not considered.
The gaugino masses are identified with the $F$-component of the
holomorphic gauge coupling, $\tau(X)$.  The $X$ dependence is originated
from the change of the beta functions due to the decoupling of a
messenger field whose mass is $X_0$ in the supersymmetric limit.
The sfermion masses can be extracted from the wave-function
renormalizations $Z_Q(X,\bar{X})$ of the matter superfields $Q$. At the
leading order, $Z_Q$ depends on $X$ through the gauge couplings
$\tau(X)$. By expanding the effective action in $F$, one can describe
the gaugino and sfermion masses by one-loop quantities,
\begin{eqnarray}
m_{\lambda}
= -F 
\left. {\partial \ln \tau(X) \over \partial X} \right|_{X = X_0}
\simeq \Delta \beta^{(1)}_{g_a} {F \over X_0},
\nonumber
\end{eqnarray}
\begin{eqnarray}
m_s^2
= - |F|^2 
\left. 
{\partial^2 \ln Z_Q (X, X^\dagger) \over \partial X \partial X^\dagger} 
 \right|_{X = X_0}
\simeq \gamma_s^{(1)} \Delta
\beta^{(1)}_{g_a}  \left| {F \over X_0} \right|^2, \label{analyticformula}
\end{eqnarray}
where $\Delta \beta^{(1)}_{g_a}$ is a discontinuity of the coefficient
of the one-loop beta function at the threshold $X_0$, and
$\gamma_s^{(1)}$ is one-loop anomalous dimension of the matter
fields. This method simplifies calculations of soft masses.

In addition to applications for gauge mediation, the technique of the
analytic continuation into superspace can be used in computing the
effective potential of the pseudo-moduli in the supersymmetry breaking
sector. In \cite{ACIII,ACIV} a leading-log effective potential for a
pseudo-moduli is computed. It turned out that, even if the potential is
generated only at higher loops, there is a regime where the potential
can be simply determined from a combination of one-loop running
data. The results were applied to survey pseudo-moduli spaces for large
classes of models \cite{ACIV}.

\subsection{General Gauge Mediation}

The method in subsection~\ref{analytic} only works in cases where the
messenger fields are weakly coupled.
In general, in order to evaluate diagrams such as in
Figures~\ref{fig:gaugino} and \ref{fig:sfermion}, we need information on
two-point functions of currents in the supersymmetry breaking dynamics.

Recently the authors of \cite{GGM} provided the most general
parametrization of the two-point functions by giving a model-independent
definition of gauge mediation: In the limit that the MSSM gauge coupling
$\alpha_i \to 0$, the MSSM sector decouples from the hidden sector that
breaks supersymmetry.
In particular, the MSSM gauge group becomes a global symmetry $G$ of the
hidden sector in this decoupling limit.
All the information we need for gauge mediation is encoded in the
currents and their correlation functions.

The conserved currents are real linear supermultiplets satisfying $D^2
{\cal J}=0$. For a $U(1)$ symmetry, the current superfield can be
written in components as
$${
{\cal J}=J+i\theta j -i \bar{\theta}\bar{j}-\theta
\sigma^{\mu}\bar{\theta}j_{\mu}+\cdots .
}$$
Two-point functions are constrained by the Lorentz invariance and the
current conservation, as follows
\begin{eqnarray*}
&&\langle J(p) J(-p) \rangle =\widetilde{C}_0(p^2/M^2),\\
&&\langle j_{\alpha}(p) \bar{j}_{\dot{\alpha}}(-p) \rangle =-\sigma^{\mu}_{\alpha \dot{\alpha}} p_{\mu} \widetilde{C}_{1/2}(p^2/M^2),\\
&&\langle j_{\mu}(p) j_{\nu}(-p) \rangle =-(p^2\eta_{\mu \nu}-p_{\mu}p_{\nu}) \widetilde{C}_1(p^2/M^2),\\
&&\langle j_{\alpha}(p) j_{\beta}(-p) \rangle =\epsilon_{\alpha \beta}M \widetilde{B}(p^2/M^2).
\end{eqnarray*}
Gaugino and scalar masses at leading order in the gauge coupling are
governed by the two-point functions\footnote{In general one-point
function $\langle J \rangle$ is non-zero, which could lead to tachyonic
sfermion. To forbid a contribution from the one-point function, the
messenger ${\bf Z}_2$ parity was imposed ${\cal J}\to -{\cal J}$ in
\cite{GGM}.  },
\begin{eqnarray}
m_{\lambda}=g^2 M \widetilde{B}(0),\qquad m_{s}^2=g^4 Y^2 A ,\label{GGMmass}
\end{eqnarray}
where $Y$ is the $U(1)$ charge of the sfermion and $A$ is the defined by
$${ A=-\int {d^4 p \over (2\pi)^4} {1\over p^2} \left(
3\widetilde{C}_1(p^2/M^2)-4 \widetilde{C}_{1/2}(p^2/M^2)+\widetilde{C}_0
(p^2/M^2) \right).  }$$ 
Note that $A$ is a linear combination of terms with different signs,
thus sfermion masses squared are not necessarily positive. Even under
this generic situation it was derived that the scalar masses obey two
sum rules, $${ {\rm Tr} (Ym^2_s)=0,\qquad {\rm Tr}((B-L)m^2_s)=0.  }$$

In generalizing the $U(1)$ theory to full Standard Model gauge group
$SU(3)\times SU(2)\times U(1)$, $A$ and $B$ are replaced by $A_k$ and
$B_k$ ($k=1,2,3$) corresponding to the three gauge group factors. These
are three real and complex numbers, providing nine independent
parameters in general gauge mediation. Arbitrary phases of $B_k$ would
typically lead to an unacceptable level of the electric dipole moment of
an electron and a neutron. So it is plausible to assume that the hidden
sector is CP invariant and to impose phenomenological constraints on the
phases. Thus we are left with six real parameters that span the
parameter space of the general gauge mediation.

It is interesting to ask if there are simple models of weakly coupled
messengers that cover the entire parameter space. This question was
first raised in \cite{CDFM} and studied in $F$-term supersymmetry
breaking models. The authors found models with the right number of
parameters, but these models have turned out not to cover the entire
parameter space. This question was examined further in \cite{GGMII},
using $D$-term supersymmetry as suggested in \cite{Nakayama}.  The
authors found models with weakly coupled messengers that span the whole
parameter space.  Though direct mediations include cases with strongly
coupled hidden sectors, the answer posed at the beginning of this
paragraph has turned out to be positive.

The idea of analytic continuation into superspace reviewed in subsection
\ref{analytic} was extended to general gauge mediation for small
supersymmetry breaking scale, $F\ll M^2$ \cite{ACI,ACII}. The authors of
\cite{ACI,ACII} showed identities for the two-point functions and
reproduced the result shown in Equation (\ref{analyticformula}) by
exploiting the general mass formulae, Equation (\ref{GGMmass}). For more
works on general gauge mediation, see
\cite{GGMIII,GGMOOPS,GGMetI,GGMetII,GGMetIII}.

\section{Direct Mediation}

With the technology of calculating or parametrizing the gaugino/sfermion
masses, we can proceed to model building with direct gauge mediation.
Direct mediation is a class of gauge mediation where messenger sectors
are responsible for supersymmetry breaking and (meta) stability of
vacua\footnote{There is also semi-direct mediation, where messengers
couple to the supersymmetry breaking sector but are not relevant for the
stability of a vacuum \cite{SeibergWecht}.  It includes mediator models
considered in \cite{AGLR}. Such models have been studied further in
\cite{SemiI,SemiII,SemiIII}.}, $e.g.$, Figures~\ref{fig:gaugino} and
\ref{fig:sfermion} as the ultimate version of it.  In some cases,
supersymmetry is restored if couplings to messengers are turned off. For
a more precise definition of direct mediation, see \cite{DM,CDFM}.
Following the seminal work \cite{DFS,DR}, this possibility of model
building has received much attention.  See
\cite{oldmodelI,oldmodelII,oldmodelIII,oldmodelIV,oldmodelV} for some of
the earlier works.

In this section, we will discuss two major challenges in constructing
phenomenologically viable models and their possible solutions: the
Landau pole and the light gaugino mass problems.

In a direct-type model, the messenger sector cannot be adjusted
arbitrarily since it is closely tied to supersymmetry breaking effects.
Especially, if there are a large number of the messenger fields, the
coupling constants of the standard model gauge interaction hit the
Landau pole below the unification scale.
One can see difficulties for example in the models of \cite{Terning,
Strassler,OzPheno}. In \cite{KOO}, a model was proposed to alleviate this issue by using a non-trivial conformal fixed point. For a more recent progress in this approach, see \cite{YanagidaConf}.

Another issue is how to generate large enough gaugino masses. Non-zero
Majorana gaugino masses require an R-symmetry to be broken\footnote{An
alternative approach is to use Dirac gaugino mass~\cite{Hall:1990hq,
Randall:1992cq, Dine:1992yw, Fox:2002bu}. This requires extra fields in
adjoint representation in the MSSM, and such models typically suffer
from the Landau pole problem. It is interesting to note that the
R-symmetry does not have to be broken in such models.  See
\cite{DiracI,DiracII} for recent proposals.}.
However, simple O'Raifeartaigh type models including the ISS model
preserve an (approximate) R-symmetry at the supersymmetry breaking
vacuum. One needs a careful arrangement to achieve supersymmetry
breaking without R-symmetry.
In addition, R-symmetry breaking is often not sufficient to guarantee
large enough gaugino masses. Below we will review these issues and
discuss how to construct successful models.

\subsection{Breaking of R-symmetry}

An R-symmetry can be broken either spontaneously or explicitly.  In
\cite{Shih}, it was shown that, for a generalized O'Raifeartaigh
model\footnote{This statement does not hold for a model with gauge
interactions \cite{DM} or a model in which two-loop Coleman-Weinberg
potential is dominant \cite{twoloop}.}, spontaneous R-symmetry breaking
requires a field with an R-charge different from $0$ or $2$.  Explicit
examples of direct mediation with exotic R-charges were constructed in
\cite{Extra}.  Since such {\it extra ordinary} gauge mediation models do
not necessarily preserve the approximate messenger parity \cite{DG96},
one-loop corrections generate tachyonic scalar masses in general. A way
to suppress such one-loop contributions was discussed in \cite{CDFM}.

Another possibility is to break an R-symmetry explicitly at the
Lagrangian level, which has an added advantage of avoiding the unwanted
R-axion.  According to \cite{NS}, for a generic superpotential, stable
supersymmetry breaking vacua require an unbroken R-symmetry. However,
explicit R-symmetry breaking is allowed if we can live with meta-stable
vacua. In particular, if R-symmetry breaking is small, supersymmetric
vacua are generated near the infinity in the field space, which
guarantees the meta-stability of the supersymmetry breaking
vacuum. Since the R-symmetry is broken, there is no symmetry reason to
prohibit the gaugino masses. However, it turns out that generating large
enough gaugino masses is still a challenge, as we will explain now.
 
\subsection{Anomalously small gaugino mass }

Let us consider the case where the supersymmetry breaking scale $\sqrt
F$ is much smaller than the messenger scale $M$.  The leading
contribution to gaugino masses in the $F/M^2$ expansion is given by
\begin{equation}
m_{\lambda} = {g^2 \over (4 \pi)^2}\cdot
F_X {\partial \over \partial X} 
\log [\, {\rm det} {\cal M}(X,M) ],\label{leading}
\end{equation}
where ${\cal M}(M)$ is fermion mass matrix of messengers and $X$ is the
pseudo-moduli field with non-zero $F$-component responsible for
supersymmetry breaking \cite{GR}. The right-hand side of this equation
vanishes in a simple O'Raifeartaigh model even if the R-symmetry
breaking is large~\cite{IzawaTobe}. This problem has been observed quite
often in models of direct gauge
mediation~\cite{DM,Terning,DirectAbel,Strassler,maru}.
Therefore, it was forced to consider models with $F/M^2 = {\cal O}(1)$,
{\it i.e.}, $\sqrt F \sim M \sim 10-100$~TeV, with which it is difficult
to avoid the Landau pole problem. Moreover, it is interesting that the models with vanishing leading order gaugino masses are severely constrained by a recent Tevatron bound on the sparticle masses and a mass bound on a light gravitino \cite{Yonekura}.

Recently, the reason for the small gaugino masses was explained in
\cite{KS} on a general ground.
Suppose that the low energy effective theory near a given vacuum is
described by the O'Raifeartaigh model with the superpotential,
\begin{equation}
W=f X +{1\over 2}(\lambda_{ab} X +m_{ab})\phi_a \phi_b+{1\over 6
 }g_{abc}\phi_a \phi_b \phi_c .
\label{eq:GO}
\end{equation}
By performing a unitary transformation of fields, one can rewrite the
general O'Raifeartaigh model in this form~\cite{Ray}.  The fields
$\phi_a$ are identified with messengers. The K\"ahler potential is
canonical for all the chiral superfields.
In this model, it is proven that $\det(\lambda X + m)$ is a constant
(independent of $X$) in a {\it stable} supersymmetry breaking vacuum,
$i.e.$, in the lowest energy state \cite{KS}.
In the proof, it is assumed that there is no unstable point anywhere in
the pseudo-moduli space since otherwise there should be a lower energy
state.

This theorem immediately means that the leading-order gaugino masses
vanish at the leading order in the $F/M^2$ expansion in this vacuum
(from Equation~(\ref{leading}) with ${\cal M} = \lambda X + m$),
regardless of whether or how an R-symmetry is broken.
Since many dynamical supersymmetry breaking models are effectively
described by Equation~(\ref{eq:GO}) near the vacuum, this puts
constraints on model building with direct gauge mediation.
The meta-stability does not help qualitatively since the presence of a
supersymmetric vacuum should be a small effect.
It is interesting that anomalously small gaugino masses is related to
global structure of the vacua of the theory and does not depend on
details on how the R-symmetry is broken.

One way to solve this gaugino mass problem is to choose our vacuum to be
even more meta-stable, $i.e.$, to use an even higher energy state
compared to the lowest supersymmetry breaking vacuum so that the
presence of an unstable point in the pseudo-moduli space is allowed.  We
will now describe such a model, which antedated the theorem in
\cite{KS}.

\subsection{Example of direct mediation model \label{sectionKOO}}

As we have discussed already, a successful model for large enough
gaugino masses requires R-symmetry breaking as well as a certain global
structure of the pseudo-moduli space. We review in this subsection the
model of \cite{KOO} as an example of models with non-vanishing gaugino
masses at the leading order in $F$.
It is a deformation of the ISS model in section~\ref{sec:susybreaking}:
an $SU(N_c)$ gauge theory with $N_f$ flavors with mass terms.

The idea is to embed the standard model gauge group into the global
symmetry of the ISS model, $SU(N_f) \times U(1)_B$ . In order to
guarantee (meta)stability, to avoid the Landau pole below the
unification scale, and to generate non-vanishing gaugino masses, it is
assumed that,

\noindent (1) the quark masses are split into two groups; $m_{\rm
light}$ for $N_c$ flavors ($Q_a$, $a=1,\cdots,N_c$) and $m_{\rm heavy}$
for $N_f - N_c$ flavors ($Q_I$, $I=1,\cdots,N_f-N_c$), and

\noindent (2) the presence of a quartic term, $(Q_I \bar Q_a)(Q_a \bar
Q_I)$, in the superpotential. The quartic term breaks the R-symmetry
explicitly. It makes the structure of the pseudo-moduli space richer. As
we will see, this helps to generate sizable gaugino masses.

\noindent 
The global symmetry is now $SU(N_f - N_c) \times SU(N_c)
\times U(1)_B$. It is shown that the Landau pole can be avoided when the
standard model gauge group is embedded to the $SU(N_f - N_c)$ part and
$m_{\rm heavy} \gg m_{\rm light}$.

In the magnetic description, the effective superpotential is given by
\begin{eqnarray}
 W = h\, \mu^2 \sum_{a=1}^{N_c} M_{aa} 
+ h\, m^2 \sum_{I=1}^{N_f-N_c} M_{II} 
- h \, \tr \, q 
M \bar q\ -h m_z \sum_{a,I} M_{Ia}M_{aI} ,
\end{eqnarray}
where $h$ is a parameter of ${\cal O}(1)$ and $ \mu \ll m$ for $m_{\rm
light} \ll m_{\rm heavy}$.
The last term (the term with a coefficient $h m_z$) corresponds to the
quartic term in the electric description.
A part of the meson fields, $M_{II}$, $M_{aI}$, and $M_{Ia}$, and of the
dual quarks, $q_I$ and $\bar q_I$, are charged under the standard model
gauge group and thus are messenger fields.
In addition to the perturbed ISS vacuum where the meson fields $M_{aa}$
acquire small VEVs which vanish as $m_z \to 0$, there are new
supersymmetry breaking vacua where the meson fields acquire large VEVs
which go to infinity as $m_z \to 0$.
The new vacua has lower energies than the perturbed ISS vacuum.

In the new vacua, gaugino masses were found to vanish at the leading
order of $F$.  We now understand the reason for it, thanks to the
theorem of \cite{KS}.

On the other hand, the original ISS vacuum has higher energy and thus
evades the theorem. Near the origin of the meson space, the low energy
effective theory is the O'Raifeartaigh model with the superpotential,
\begin{eqnarray}
W=h \tr \left(
\mu^2 X
- X\rho \tilde{\rho}
- m e^{-\theta}\rho \tilde{Z}
- m e^{\theta}Z \tilde{\rho}
- m_zZ\tilde{Z}  \right). \nonumber 
\end{eqnarray}
One-loop effects stabilize the potential, giving rise to the ISS
vacuum. However, the tree level scalar potential has an unstable region
near the new vacua, $X \sim m^2 / m_z$.
It is because of this that the ISS vacuum evades the theorem of
\cite{KS} and generates the gaugino masses,
\begin{eqnarray}
 m_\lambda = \frac{ g^2 N_c }{(4 \pi)^2} \frac{h  \mu^2}{
  m} \frac{ m_z }{  m}
+ O\left( \frac{m_z^2}{ m^2} \right)\ ,
\end{eqnarray}
where $g$ is the coupling constant of the standard model gauge
interaction. Scalar masses are also obtained by two-loop diagrams and
can be adjusted to be the same order as the gaugino mass ${\cal
O}(m_i)\simeq {\cal O}(m_{\lambda})$ by setting $m_z \sim m / \sqrt{
N_c}$.

\section{Stringy Realization}

We are going to have an interlude to describe realization of
supersymmetry breaking mechanisms and their mediations in string theory.
There are two motivations for stringy constructions. One is to provide a
unified framework for field theory models to understand the nature of
the parameters of these models.  Moreover, string theory has been used
to develop powerful computational tools for field theory effects and to
gain geometric insights into supersymmetry breaking phenomena.  Another
motivation is to understand supersymmetry breaking in string theory
better. In the past quarter-century, much progress has been made in
understanding of string theory in supersymmetric backgrounds and many
exact results on non-perturbative effects have been derived. It is
desirable to extend these results to non-supersymmetric situations.

Once we accept meta-stable vacua, a wide variety of stringy realizations
becomes possible.  Since long-lived meta-stable vacua are ubiquitous in
vector-like models such as SQCD, it is relatively easy to embed
supersymmetry breaking models in string theories. In this section, we
will review model building by intersecting branes, gauge/gravity
dualities and geometric transitions. For an interesting recent
development of string phenomenology with $F$-theory, see \cite{Ftheory}
for a review.

\subsection{Brane configuration}

Open strings are collective coordinates of D branes. On parallel D$p$
branes, if we take the limit of string length $\ell_s \rightarrow 0$
while scaling the string coupling constant as $g_s = g_{{\rm YM}}^2
\ell_s^{3-p}$, the low energy effective theory becomes the maximally
supersymmetric Yang-Mills theory in $(p+1)$ dimensions with the gauge
coupling given by $g_{{\rm YM}}$.  We can reduce the number of
supercharges by considering configurations of intersecting branes. For
example, if we have a pair of parallel NS 5-branes, and if we suspend
parallel D4 branes between them in such a way that all the branes share
4 dimensions, then the low energy effective theory is the ${\cal N}=2$
supersymmetric pure Yang-Mills theory in 4 dimensions.  Preserving
supersymmetry requires that branes be in specific relative angles. Any
other angle would break supersymmetry completely. See \cite{GKreview}
for a review.

\begin{figure}
\centerline{\epsfbox{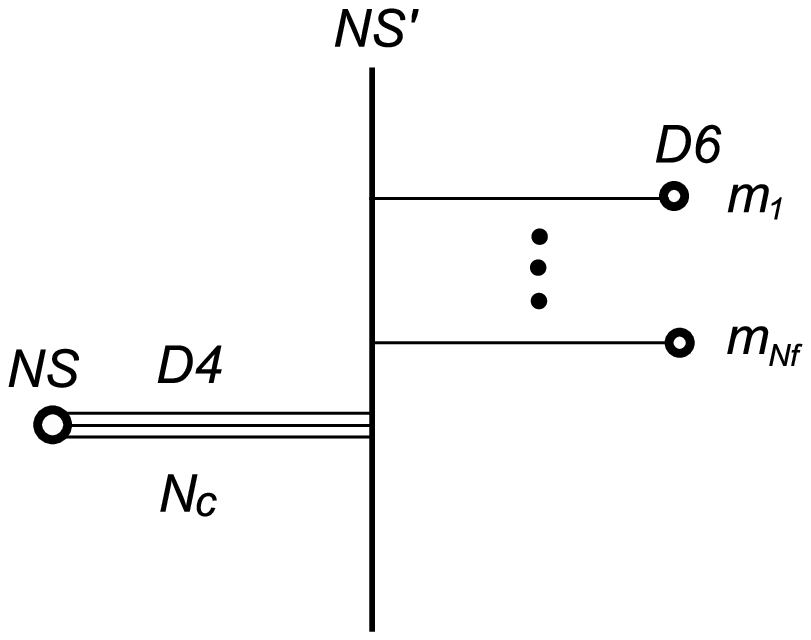}} 
\caption{The electric brane configuration for ${\cal N}=1$ supersymmetric gauge theory with massive flavors.}  \label{Massive}
\end{figure}

\begin{figure}
\centerline{\epsfbox{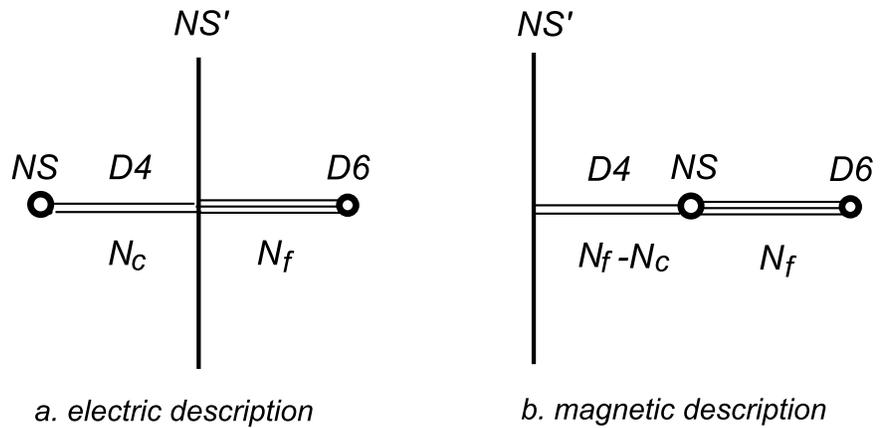}} 
\caption{The supersymmetric brane configurations for ${\cal N}=1$ supersymmetric gauge theory with massless flavors. }  \label{Massless}
\end{figure}

Stringy description of the ISS model and its meta-stable vacua was
initiated in \cite{BraneI,BraneII} and studied further in
\cite{BraneIII,BraneIV}.  Figure (\ref{Massive}) shows the brane
configuration for ${\cal N}=1$ $SU(N_c)$ gauge theory with $N_f$
fundamental chiral multiplets with masses $m_1,...,m_{N_f}$, which is
the ISS model. The meta-stable supersymmetry breaking vacua of the ISS
model can be found by going to the magnetic description.  To go from the
electric description to the magnetic dual, we exchange the location of
the two NS 5-branes.  The number of branes changes during this process
due to the Hanany-Witten mechanism. Figures (\ref{Massless}.a) and
(\ref{Massless}.b) show the electric and magnetic branes configurations
when masses are equal to zero. We see that the gauge group in the
magnetic description is $SU(N_f-N_c)$ as expected by the Seiberg
duality.  In both description, D4 branes are parallel and supersymmetry
is preserved.

\begin{figure}
\centerline{\epsfbox{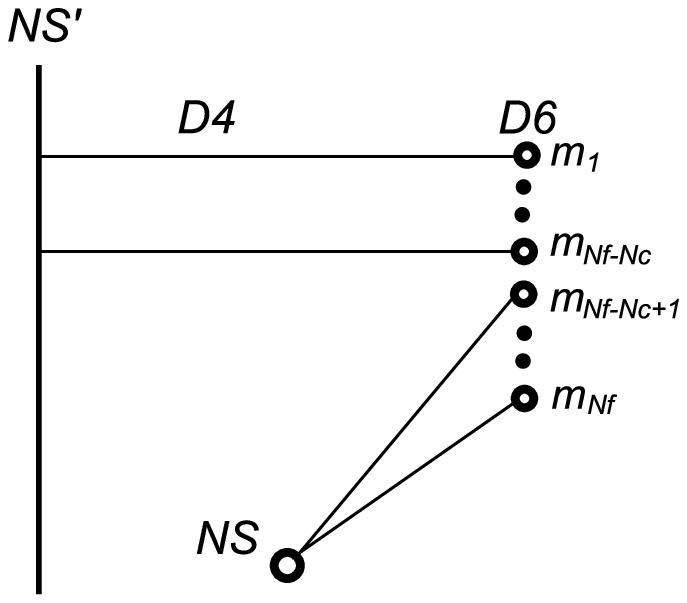}} 
\caption{The brane configuration for the meta-stable supersymmetry breaking vacuum. }  \label{SUSYbreaking}
\end{figure}

Now, let us turn masses to non-zero. In the electric description, this
can be done by moving the D6 branes up an down as in Figure
(\ref{Massive}).  In the magnetic description, however, we cannot move
the D6 branes in this way while keeping all the D4 branes parallel to
each other. We can keep $(N_f-N_c)$ D4 branes in parallel by
reconnecting them at the NS 5-branes, but the remaining $N_c$ D4 branes
have to be tilted as shown in Figure (\ref{SUSYbreaking}). The resulting
configuration breaks all the supersymmetry.

The brane configuration in this limit reproduces various features of
vacuum structure of the ISS model such as global symmetries,
pseudo-moduli and vacuum energy, as explained in
\cite{BraneI,BraneII}. Thus, it is reasonable to identify this
configuration with the supersymmetry breaking meta-stable vacuum in the
field theory.

In \cite{BraneIII}, it was claimed that this brane configuration is not
related to the original electric description and that it does not
describe the ISS vacuum because of a certain asymptotic behavior of D
branes caused by brane bending at finite $g_s$.  In \cite{BraneIV}, an
alternate interpretation of the brane bending involving string tachyons
was proposed and it was argued that the brane configuration of
\cite{BraneI,BraneII} indeed describes the ISS vacuum. We refer to these
papers for details on the issue.

\subsection{Holographic Gauge Mediation}

Soft-terms in gauge mediation models are related to correlators of
global symmetry currents in the hidden sector \cite{GGM}, as we reviewed
in section 4.2. Such relations are particularly useful for direct-type
models where hidden sectors are strongly coupled but the current
correlators may be calculable.  In \cite{GGMOOPS}, this idea was applied
to a class of strongly coupled hidden sectors to derive low energy
parameters of the models.  Gauge/gravity duality in string theory can be
useful since the current correlators in field theories are computed by
Green's functions of the corresponding gauge fields in the gravity
duals.

One of the well-studied examples is the duality between the cascading
${\cal N}=1$ $SU(N+M)\times SU(N)$ gauge theory and the warped deformed
conifold geometry in type IIB string theory \cite{KlebanovStrassler}.
In \cite{KPV}, a meta-stable supersymmetry breaking vacuum was
constructed in this setup by putting anti-D3 branes at the tip of this
conifold geometry. This vacuum can decay into a supersymmetric vacuum by
quantum tunneling and brane/flux annihilation. Since the process happens
near the tip of conifold, the UV description is not modified and we
expect that there is a field theory dual to describe the process.  Both
the supersymmetric vacuum and the meta-stable vacuum must have
corresponding states in such a field theory. Recently, this construction
was generalized in \cite{ABFKI,ABFKII} to quiver gauge theories and
string theory on quotients of the conifold, and the correspondence
between meta-stable vacua in the field theories and in the gravity
theories have been clarified. In particular, the model studied in
\cite{ABFKII} gives a stringy realization of the direct gauge mediation
model constructed in \cite{KOO} and discussed in the previous
section. Moreover, it provides a natural mechanism to generate small
parameters in the model.

In the holographic gauge mediation model\footnote{For earlier related
works in five dimensional theory in a truncated $AdS_5$ background, see
\cite{Gherghetta:2000qt,Gherghetta:2000kr,NomuraI,NomuraII,NomuraIII}.}
of \cite{HolographicMediation}, strongly coupled messengers and hidden
sectors are replaced by the supersymmetry breaking solutions of Type IIB
supergravity constructed in \cite{DKM}, which takes into account the
back reaction of the anti-branes.  Although gaugino condensation breaks
the R-symmetry to ${\bf Z}_2$, the issue of anomalously small gaugino
masses is not resolved in this model. This construction was generalized
in \cite{MSSII}, by using the supersymmetry breaking solution of
\cite{MSSI}.

\subsection{Geometric Transition}

Another way to construct low energy gauge theories is to use D branes
wrapping cycles of Calabi-Yau manifolds. For example, consider a
resolved conifold and wrap its small 2-sphere with $N$ D5 branes. The
low energy effective theory is the ${\cal N}=1$ pure Yang-Mills theory
in 4 dimensions with gauge group $U(N)$.  By a chain of dualities, one
can relate this construction to an intersecting brane configuration of
the type discussed earlier in this section \cite{OVbrane}.

The geometric transition relates the resolved conifold to a deformed
conifold with a small 3-sphere. In particular, the
Veneziano-Yankielowicz superpotential of the ${\cal N}=1$ Yang-Mills
theory realized on the $N$ D5 branes on the resolved conifold is equal
to the Gukov-Vafa-Witten superpotential for the deformed conifold with
$N$ units of Ramond-Ramond fluxes through the 3-sphere \cite{Vafa}. This
is an example of the large $N$ duality and is closely related to the
gauge/gravity duality of \cite{KlebanovStrassler,MaldacenaNunez}. It has
been generalized to a variety of gauge theories and corresponding
geometries, and their superpotentials are computed exactly using
topological string theory \cite{CIV}.

In \cite{Vafa}, it was suggested that the geometric transition can be
used even when both branes and anti-branes are present.  This proposal
was made more precise in \cite{VafaI,VafaII,ABF}.  It was conjectured
that the large $N$ limit of brane/anti-brane systems are Calabi-Yau
manifolds with fluxes and that the physical potential can be computed
using the method of topological string theory as in supersymmetric
cases, except that some of the fluxes are negative.

In \cite{ABF,BMV}, field theory descriptions of such meta-stable vacua
are given. Their M-theory realizations have been explored in
\cite{MasakiI,MasakiII}.

\section{Electroweak symmetry breaking}

The final topic is the most relevant for particle physics. In the MSSM,
electroweak symmetry breaking does not happen in the supersymmetric
limit since supersymmetry guarantees the absence of the negative mass
squared for the Higgs boson. Thus, there must be a coupling between the
supersymmetry breaking dynamics and the Higgs sector of the MSSM.

The use of a technicolor dynamics to break both supersymmetry and
electroweak symmetry was proposed in~\cite{DSBWitten,DFS,DR}.  The Higgs
fields obtain VEVs through a direct coupling to the dynamics, and give
masses to fermions in the Standard Model.
On the other hand, in calculable models such as in
\cite{DNY,DineNelsonShirman,DineNelson}, one usually assumes that the
Higgs sector is separated from the supersymmetry breaking dynamics and
some communication through messenger fields generates the Higgs
potential and the Higgsino mass.

There are, in fact, number of difficulties in this program and
completely successful models have not been found. Here we explain the
difficulties and current status. First, we list the phenomenological
requirements:
\begin{enumerate}
 \item The $Z$ boson mass, $m_Z = 91$~GeV,
 \item The Higgs boson mass constraint, $m_h > 114$~GeV~\cite{Barate:2003sz},
 \item The Higgsino mass constraint, $m_{\tilde H} > 94$~GeV~\cite{Abdallah:2003xe},
 \item The gaugino mass constraint, $m_{\lambda} >
       O(100)$~GeV~\cite{Abdallah:2003xe}\footnote{One of the
       neutralinos can be very light~\cite{Hooper:2002nq,
       Belanger:2002nr, Dreiner:2009ic}.},
 \item The top-quark mass, $m_t = 173$~GeV~\cite{top:2009ec},
 \item The bound on the electric dipole moments of the neutron, $d_n < 3
       \times 10^{-26} e$~cm~\cite{Baker:2006ts}.
\end{enumerate}
As we will explain below, it is not easy to naturally explain all of the
above in supersymmetric models. This is generally called the
$\mu$-problem.
Most of the discussion below are not specific to gauge mediation models,
but the problems are particularly sharp in gauge mediation since it is
designed to be calculable.

The Higgs potential in the MSSM is 
\begin{eqnarray}
 V &=& (m_{H_u}^2 + \mu^2 ) |H_u|^2 + (m_{H_d}^2 + \mu^2 ) |H_d|^2
+ \left( B \mu H_u H_d + {\rm h.c.} \right)
\nonumber \\
&&+ {1 \over 8} (g_Y^2 + g_2^2) (|H_u|^2 - |H_d|^2)^2,
\label{eq:higgs}
\end{eqnarray}
where $H_u$ and $H_d$ are the neutral components of the Higgs fields in
the MSSM. The part of the quadratic terms $\mu^2$ is supersymmetric
contribution to the Higgs potential. By definition, $\mu$ is the mass of
the Higgsino. By supersymmetry, the quartic potential is related to the
gauge couplings, $g_Y$ and $g_2$, which control the Higgs boson
mass. Other quadratic terms are soft supersymmetry breaking terms which
should arise from couplings to the supersymmetry breaking sector. The
$B$ parameter is in general complex valued.

If the Higgs fields are only weakly coupled to the supersymmetry
breaking sector such as in \cite{DFS}, the typical diagram to generate
the Higgsino mass (the $\mu$-term) is given in Figure~\ref{fig:higgsino}
where the technicolor box is replaced by a messenger loop or a
supersymmetry breaking box.
The problem is that if there is a one-loop diagram as in
Figure~\ref{fig:higgsino}, {\it one-loop} diagrams for generating
$m_{H_u}^2$, $m_{H_d}^2$ and $B \mu$ should also be present.
Since $\mu^2$ is effectively {\it two-loop} valued, it suggests that the
typical mass scale for the Higgs potential is larger than the Higgsino
mass by a one-loop factor, $i.e.$, $m_Z \gg \mu$. Now we can see
inconsistency with the items 1 and 3.

One could have evaded the above problem by assuming that the Higgs
fields do not acquire VEVs from their potential and there are other
sources for the $Z$ boson mass such as technicolor models in
\cite{DFS,DR}. Small VEVs can be obtained through the diagram in the
right-hand-side of Figure~\ref{fig:techni}, and they are responsible for
the fermion masses.
However, since the top quark is rather heavy (item 5), such an
assumption is not easily realized in a consistent way.

Even if we somehow avoid the hierarchies in the parameters in the Higgs
potential at the messenger scale dynamics, there are also one-loop
corrections to $m_{H_u}^2$ and $m_{H_d}^2$ which are proportional to the
gaugino masses squared. Therefore, the gaugino masses cannot be very
large compared to the $Z$-boson mass.
Together with the item 4, the gaugino masses are ${\cal O}(100)$~GeV.
In gauge mediation, the gaugino masses are obtained at one-loop
level. Therefore, it suggests that the Higgs fields are also weakly
coupled to the supersymmetry breaking sector. In this case, the
hierarchy mentioned above among $m_{H_u}^2$, $m_{H_d}^2$, $B \mu$ and
$\mu^2$ is a quite generic consequence.

In the Higgs potential in Equation~(\ref{eq:higgs}), there is a
tree-level prediction for the Higgs boson mass: $m_h \leq m_Z$, which is
clearly inconsistent with the items 1 and 2. This requires a rather
large quantum correction to the quartic couplings arising from
supersymmetry breaking~\cite{Okada:1990vk, Ellis:1990nz,
Haber:1990aw}. In the MSSM, the largest contribution is from the
top-stop loop diagrams. However, if there is a large quantum correction
to the quartic coupling constant, there is also a large contribution to
the quadratic term, especially to $m_{H_u}^2$. Such a large contribution
calls for fine-tuning of the parameters to have a correct size of the
electroweak VEV, $i.e.$, the $Z$-boson mass~\cite{Barbieri:1987fn} (See
\cite{Chacko:2005ra, Kitano:2006gv} for recent discussions.).

The degree of fine-tuning gets relatively milder when we have a sizable
stop-stop-Higgs coupling called the $A$-term. However, a simple gauge
mediation model predicts $A=0$ at the messenger scale with which ${\cal
O}(1\%)$ tuning is necessary~\cite{Kitano:2006gv}.

One can try to build a model to generate the $A$-term. However, such a
model requires that the $A$-term has the same phase as the gaugino
masses, since the relative phase is physical and observable.  The
constraint from item 6 restricts the phase to be smaller than ${\cal
O}(10^{-(2-3)})$. The same restriction applies for the $B$ parameter in
the Higgs potential.

Several ways to address those problems have been considered in the
literature.
As recent progress in gauge mediation, for example, semi-strongly
coupled models are considered in \cite{SweetspotII} to avoid large
hierarchies among $m_{H_u}^2$, $m_{H_d}^2$, $B \mu$ and $\mu^2$, where
the CP phase is controlled by a symmetry.
A perturbative model without the hierarchy is presented
in~\cite{Giudice:2007ca} where the Higgs and the messenger sectors are
extended. (See also \cite{DNY, muI, Agashe:1997kn, deGouvea:1997cx} for
earlier discussions.)
In \cite{Csaki:2008sr}, it has been pointed out that the correct size of
$m_Z$ can be obtained even in the presence of the hierarchy.  An
explicit example to realize the correct hierarchy pattern was presented.
A mechanism to suppress $m_{H_u}^2$, $m_{H_d}^2$ and $B \mu$ relative to
$\mu^2$ by large renormalization effects has been discussed in
\cite{Roy:2007nz, Murayama:2007ge}.
In \cite{GGMIII}, discussions were reformulated in the context of
general gauge mediation. For another recent approach using more than one supersymmetry breaking spurion, see \cite{Liu}.

One should not forget that natural electroweak symmetry breaking is one
of the main motivations to consider the supersymmetric standard model.
Supersymmetry breaking and its connection to the Higgs sector should be
arranged so that electroweak symmetry breaking happens naturally.
We encountered problems in this connection, but, on the other hand, the
need for a special structure of the Higgs sector may be a hint for the
actual underlying theory.
It is desirable to have more ideas to be confirmed in future or on-going
experiments, such as the LHC experiments, in order to reveal the role of
supersymmetry in Nature.

\section*{Acknowledgments}

We would like to thank M. Graesser, Z. Komargodski, and D. Shih for
comments on this manuscript. RK is supported in part by the Grant-in-Aid
for Scientific Research 21840006 of JSPS.  HO is supported in part by US
DOE grant DE-FG03-92-ER40701, the World Premier International Research
Center Initiative of MEXT of Japan, and by a Grant-in-Aid for Scientific
Research (C) 20540256 of JSPS.  YO's research at Perimeter Institute for
Theoretical Physics is supported in part by the Government of Canada
through NSERC and by the Province of Ontario through MRI.

\section{NUMBERED LITERATURE CITED}

\bibliographystyle{arnuke_revised.bst}
\bibliography{susybreakingREF.bib}

 \end{document}